# Prediction of $BaBiO_3$-like superconducting perovskites in K-doped $SrAsO_3$

Zhihong Yuan[1,2], Rui Liu[1], Pengyu Zheng[1], and Zhiping Yin[1,3*]

[1] School of Physics & Astronomy and Center for Advanced Quantum Studies, Beijing Normal University, Beijing 100875, China

[2] School of Physics and Electronic Engineering, Shanxi Normal University, Taiyuan 030031, China

[3] Key Laboratory of Multiscale Spin Physics (Beijing Normal University), Ministry of Education, Beijing 100875, China

Using first-principles calculations, we predict a new perovskite compound $SrAsO_3$. The undoped cubic phase has pronounced soft-phonon instabilities, which are gradually suppressed upon K doping the Sr site. The cubic phase becomes dynamically stable for K-doping levels above approximately 60%, and the stabilized K-doped phases are metallic with predicted conventional phonon mediated superconductivity. Moreover, the inclusion of nonlocal exchange interactions broadens the electronic bandwidth, enhances the electron-phonon coupling (EPC) strength, and increases the superconducting transition temperature ($T_c$) of these doped compounds. In particular, the HSE06 hybrid exchange-correlation functional corrected EPC constant $\lambda$ reaches 1.41 for $Sr_{0.4}K_{0.6}AsO_3$, corresponding to a predicted $T_c$ of 44.3 K. These results suggest that $SrAsO_3$ is a $BaBiO_3$-like superconducting perovskite driven by strong electron-phonon coupling.

## I. INTRODUCTION

$ABO_3$-type perovskite superconductors constitute a unique branch in the field of superconductivity[1-5]. Although their $T_c$ values are lower than those of high-temperature superconductors such as copper- and iron-based systems, their relatively simple cubic structure, absence of magnetic ions and strong EPC make them important systems for understanding phonon-mediated superconductivity in oxide materials.

As early as 1964, superconductivity was discovered in reduced strontium titanate ($SrTiO_3$), a perovskite oxide, albeit with a very low $T_c$ of about 0.4 K[2]. In 1975, superconductivity was reported in the $BaPb_{1-x}Bi_xO_3$ (BPBO), with a $T_c$ of approximately 13 K near $x = 0.25$ [3]. Related Pb-based perovskites were also explored by substituting Sb for Pb, and superconductivity was reported in $BaPb_{0.75}Sb_{0.25}O_3$, with a lower $T_c$ of about 3.5 K[6]. A major breakthrough occurred in 1988 with the successful synthesis of $Ba_{1-x}K_xBiO_3$ (BKBO) [4]. This system exhibits a significantly enhanced $T_c$ of about 30 K at an optimal doping level of $x \approx 0.4$, representing the highest $T_c$ reported among non-cuprate oxide superconductors at that time. BKBO is the most representative and widely studied $ABO_3$-type perovskite superconductor. The parent compound $BaBiO_3$ is a charge density wave (CDW) insulator, in which the insulating state is closely associated with Bi-O breathing distortions and Bi charge disproportionation. Upon K substitution on the Ba site, the CDW-related lattice distortion is suppressed, the system becomes metallic, and superconductivity emerges. Therefore, BKBO provides a prototypical example in which superconductivity is closely linked to the suppression of lattice instabilities and strong EPC. In recent years, both theoretical and experimental studies have further shown that nonlocal correlation effects are essential for accurately describing its electronic structure, EPC strength, and relatively high $T_c$[7-9].

Following the discovery of BKBO, considerable efforts have been devoted to expanding the family of Bi-based $ABO_3$ perovskite superconductors. In 1997, S. Kazakov *et al*. reported a second family of Bi-based perovskite superconductors derived from $SrBiO_3$, where K substitution induces superconductivity in $Sr_{1-x}K_xBiO_3$ over the composition range $x$ = 0.45-0.60, with $T_c$ of about 12 K. They also reported superconductivity in $Sr_{0.5}Rb_{0.5}BiO_3$, with $T_c$ of about 13 K[5]. In 1998, N. Khasanova *et al*. discovered superconductivity in $(K_{0.87}Bi_{0.13})BiO_3$ with a $T_c$ of approximately 10 K[10]. Subsequent studies further expanded the compositional range of superconducting bismuthates, indicating that carrier doping and structural stabilization are both important for superconductivity in Bi-based perovskites[11-13].

More recently, the discovery of $Ba_{1-x}K_xSbO_3$ (BKSO) extended the search for $ABO_3$-type superconductivity from Bi to Sb. In 2021, Kim *et al*. successfully stabilized BKSO for the first time and reported a $T_c$ of approximately 15 K at the optimal doping level of $x \approx 0.65$[14]. Subsequently, we performed calculations and found that the BKSO are extraordinary BCS superconductors, where non-local electronic correlation is also crucial to generate the required strong EPC[15].

Motivated by the superconductivity observed in these $BaBiO_3$-like superconductors, whether other $ABO_3$-type perovskite compounds can exhibit superconductivity remains an open question. This work aims to employ first-principles calculations to predict novel $BaBiO_3$-like cubic perovskite superconductors and investigate their superconducting properties. Considering that arsenic (As) is isovalent with bismuth (Bi) and antimony (Sb) (belonging to the same group of the periodic table), the perovskite $SrAsO_3$ and $BaAsO_3$ are expected to exhibit band structures similar to those of $BaBiO_3$ and $BaSbO_3$ near the Fermi level. This similarity suggests that optimal doping may induce superconductivity, with the isotope effect favoring a higher $T_c$ in $(Sr,K)AsO_3$ (SKAO) and $(Ba,K)AsO_3$ (BKAO). The structures, electronic structures, lattice dynamics, electron-phonon coupling, and superconductivity of SKAO and BKAO are systematically investigated.

The rest of the paper is organized as following: In Section II, the computational methods in this work are described. Section III reports the crystal structure, electronic structure, lattice dynamics, EPC and $T_c$ of SKAO. Section IV presents calculated results of BKAO. In Section V, a summary of this work is given.

## II. METHODS

Since no crystallographic information for $SrAsO_3$ is available in the ICSD (Inorganic Crystal Structure Database), the initial cubic perovskite structure was constructed from the ideal cubic $BaBiO_3$ structure through elemental substitution, followed by full structural optimization within the density functional theory (DFT) local-density approximation (LDA)[16] framework. After establishing the parent structure, a series of K-doped crystal structures were generated within the virtual crystal approximation (VCA) method[17].

The electronic band structures and densities of states (DOS) of the doped systems were calculated using both the DFT-LDA and DFT-HSE06 (Heyd-Scuseria-Ernzerhof hybrid functional)[18] methods, as implemented in the VASP[19]. The lattice dynamical properties and EPC were investigated within density functional perturbation theory (DFPT)[20] using the LDA exchange-correlation functional, as implemented in Quantum ESPRESSO[21]. Norm-conserving pseudopotentials[22] were employed, with kinetic-energy and charge-density cutoffs of 100 Ry and 400 Ry, respectively. For the phonon and EPC calculations, $k$-point meshes of $24 \times 24 \times 24$ and $q$-point meshes of $6 \times 6 \times 6$ were used, respectively. The EPC strength was further renormalized to include nonlocal exchange effects within the HSE06 hybrid functional, following Ref.[7].

## III. RESULTS

### 3.1 Structural and Stability of $SrAsO_3$

Figure 1 shows the schematic cubic perovskite structure of $SrAsO_3$. The optimized lattice constant is 3.879 Å. To evaluate the energetic stability of the cubic phase, we considered a possible synthesis reaction, $SrAs_2O_6 + Sr = 2SrAsO_3 + Q$, where Q denotes the energy released during the reaction. All reactants were taken from experimentally synthesized phases available in the ICSD database. The calculated reaction energy is approximately 1.69 eV per $SrAsO_3$ unit, indicating that the formation of $SrAsO_3$ is energetically favorable and suggesting its potential experimental accessibility.

The dynamic stability of $SrAsO_3$ was further examined using the DFPT. As shown in Fig. S1(a), the cubic parent phase exhibits pronounced imaginary frequency phonon modes, indicating strong dynamical instability. Analysis of the phonon eigenvectors shows that the $\Gamma$-point instability is dominated by a Slater-type polar soft mode, primarily involving displacement of the As atom against the oxygen octahedron, with a minor contribution from the Sr atom. In contrast, the zone-boundary instabilities are nonpolar and mainly associated with As-O bond distortions. In particular, the $M$-point instability is dominated by As-O bond stretching modes (Fig 1(b)), while the $R$-point instability corresponds to a breathing mode of the $AsO_6$ octahedral (Fig 1(c)). These modes are similar to the soft phonon instabilities observed in $BaBiO_3$ and $BaSbO_3$.

### 3.2 Stability and Electronic structures of SKAO

In $BaBiO_3$ and $BaSbO_3$ perovskites, K doping is known to suppress the CDW related Bi-O (Sb-O) breathing distortion and induce superconductivity. Motivated by this behavior, we investigated whether K doping can similarly suppress the As-O bond distortion related instability in the present system. Accordingly, phonon spectra were calculated at different K-doping levels to assess the dynamical stability of the cubic phases. As shown in Fig. S1, the imaginary phonon branches gradually weaken with increasing K doping and disappear completely for $x \geq 0.5$, indicating that K doping stabilizes the cubic perovskite phase. Based on the above dynamical stability analysis, we further investigated the electronic structures, EPC, and superconducting properties of $Sr_{1-x}K_xAsO_3$ ($x$ = 0.5, 0.6, 0.7, 0.8, 0.9, 1.0). All these doped compounds are found to be metallic and are predicted to exhibit phonon-mediated superconductivity.

The electronic band structures calculated using both the LDA and HSE06 functionals are presented in Fig. 2. For all doping levels, the HSE06 functional yields significantly broader band widths compared with LDA, as summarized in Tables S1 (*Γ-X* and *Γ-M* directions). The corresponding band-broadening factor remains nearly unchanged with increasing K doping, indicating that the enhancement of the band width induced by the HSE06 hybrid functional is essentially independent of the doping level. As shown in Fig. 3, the HSE06 functional shifts the valence-band maximum to lower energies relative to LDA. However, the magnitude of this shift decreases with increasing K doping and becomes negligible at $x$ = 1.0 ($KAsO_3$), where the valence-band maxima from the two functionals nearly coincide. In general, the HSE06-induced shift decreases as the valence-band maximum approaches the Fermi level. Overall, the HSE06 hybrid functional not only broadens the conduction bands but also modifies the position of the valence bands across all doped systems.

To analyze the orbital contributions to the electronic structure, Fig. 4 presents the DFT-HSE06 orbital-resolved band structure together with the total and partial density of states (DOS) for $Sr_{0.4}K_{0.6}AsO_3$ as a representative case. The states near the Fermi level are mainly derived from O-2$p$ and As-4$s$ orbitals. This is consistent with the partial DOS, where the energy window from -2.5 to 4.4 eV is dominated by O-2$p$ and As-4$s$ orbitals contributions.

### 3.3 Lattice dynamics and EPC of SKAO

We now present the main results of lattice dynamical and EPC properties of $Sr_{1-x}K_xAsO_3$ ($x$ = 0.5, 0.6, 0.7, 0.8, 0.9, 1.0) obtained from DFPT-LDA calculations. The calculated phonon dispersion curves, the Eliashberg phonon spectral function $\alpha^2F(\omega)$ and the cumulative frequency dependence of EPC strength $\lambda(\omega)$ are shown in Fig. 5, where the size of red dots superimposed on the phonon dispersion curves is proportional to the mode- and momentum-dependent EPC strength $\lambda_{qv}$. The projected phonon density of states (right panel in Fig 5) shows that the high-frequency modes are dominated by oxygen vibrations, whereas low- and intermediate-frequency regions involve mixed vibrations of the Sr, As, and O atoms. With increasing K doping, the phonon spectrum shifts toward higher frequencies, and the high-frequency phonon DOS becomes progressively broader.

The mode-resolved EPC ($\lambda_{qv}$) in Fig. 5(a)-(c) shows that both the oxygen-oscillating mode at the $X$ point (Fig 1(a)) and the oxygen-stretching mode at the $M$ point (Fig 1(b)) provide significant contributions to the EPC for $x$ = 0.5-0.7. As the K doping level increases, the $M$ point oxygen-stretching mode progressively hardens, resulting in a continuous suppression of its EPC contribution, which becomes negligible at $x$ = 0.8. Conversely, decreasing the K doping level from $x$ = 0.8 to $x$ = 0.5, the oxygen-stretching mode at the $M$ point becomes progressively softened, accompanied by an enhancement of its EPC strength [Fig. 5(d)–(a)]. When $x < 0.5$, this mode becomes dynamically unstable with imaginary frequencies, indicating a lattice instability of the cubic phase and a tendency toward a CDW-like distortion. The evolution of $\alpha^2F(\omega)$ and $\lambda(\omega)$ further confirms that the total EPC strength increases as the K doping level decreases from 0.8 to 0.5. At higher doping levels, $x$ = 0.9 and 1.0 [Fig. 5(e)–(f)], the oxygen-oscillating mode at the $X$ point remains important, while additional contributions from low-frequency oxygen-rotational modes at the $M$ and $R$ points (Fig 1(d)) become significant. As a result, the $\alpha^2F(\omega)$ is redistributed toward lower frequencies, leading to a pronounced increase of $\lambda(\omega)$ in the low-frequency region.

By integrating $\alpha^2F(\omega)$ for $Sr_{1-x}K_xAsO_3$ ($x$ = 0.5-1.0), the total EPC constant $\lambda$ and the logarithmic average phonon frequency $\omega_{\log}$ were obtained. Using these DFPT-LDA values together with the Allen-Dynes formula and $\mu^*$ = 0.1, the $T_c$ was estimated, as summarized in Table 1 and Fig 6. Within the DFPT-LDA, $T_c$ for the SKAO systems reach 36.1 K (with $\lambda \approx 1.47$) at a K doping level of $x$ = 0.5.

### 3.4 Realistic EPC and $T_c$ of SKAO

The results discussed in Section 3.2 indicate that nonlocal exchange interactions have a pronounced impact on the electronic structures of SKAO. In this section, we further evaluate their influence on the EPC strength and $T_c$ by following the approach of Ref. [7], in which DFPT-LDA calculations are combined with HSE06 supercell frozen-phonon calculations to estimate the realistic EPC strength.

To quantify the EPC beyond the DFPT-LDA level, we calculated the reduced electron-phonon matrix elements (REPMEs) for selected strongly coupled phonon modes. Before applying this correction, we first examined the stability of the cubic phase against $M$ point oxygen-stretching distortions using the HSE06 functional. At $x$ = 0.5, the oxygen-stretching distorted supercell is lower in total energy than the undistorted cubic supercell, indicating that the cubic phase remains unstable with respect to this distortion at this doping level. In contrast, at $x$ = 0.6, the undistorted cubic supercell becomes energetically more favorable. These results suggest that, within the HSE06 functional, the oxygen-stretching instability is suppressed for $x \geq 0.6$, consistent with the stabilization of the ideal cubic perovskite phase.

Based on the DFPT-LDA mode-resolved EPC ($\lambda_{qv}$) results shown in Fig 5, representative strongly coupled phonon modes were selected for the REPME calculations. For $x$ = 0.6 and 0.7, REPMEs were evaluated for both the oxygen-oscillating mode at the $X$ point and the oxygen-stretching mode at the $M$ point. For $x$ = 0.8, only the oxygen-oscillating mode at $X$ point was considered. For $x$ = 0.9 and 1.0, REPMEs were evaluated for the oxygen-oscillating mode at the $X$ point as well as the oxygen-rotational modes at the $M$ and $R$ points. The calculated band splittings and the corresponding REPME values are summarized in Tables S2 and illustrated in Figs. S2-S6. Based on these REPMEs, enhancement factors for the EPC strength were estimated within the HSE06 functional. Combining these enhancement factors with the DFPT-LDA results, we obtained the corrected EPC strength $\lambda_H$, logarithmic average phonon frequency $\omega_{\log,H}$ and $T_c$ for the considered doping levels, as summarized in Table 1 and Fig 6. The results indicate that nonlocal exchange interactions significantly enhance both $\lambda$ and $T_c$ across all doped compounds. After incorporating nonlocal exchange effects, the $T_c$ of $Sr_{0.4}K_{0.6}AsO_3$ increase to approximately 44.3 K, with a corresponding $\lambda$ value of 1.41.

## IV. DISCUSSION

In addition to the $SrAsO_3$ discussed above, we also investigated the cubic perovskite $BaAsO_3$. After structural optimization, the equilibrium lattice constant of the cubic phase was determined to be 3.972 Å. A possible synthesis reaction was also constructed for this compound $2BaO + As_2O_4 = 2BaAsO_3 + Q$, yielding a reaction energy of approximately 1.37 eV per $BaAsO_3$ unit, which suggests that the predicted cubic perovskite $BaAsO_3$ may be experimental accessible. DFPT-LDA phonon calculations show that the undoped cubic phase $BaAsO_3$ exhibits pronounced imaginary phonon modes, indicating a dynamical instability. Similar to $SrAsO_3$, K doping effectively suppresses these lattice instabilities. At a K-doping level of $x = 0.7$, all imaginary phonon modes disappear (Fig. S7), indicating that the cubic phase becomes dynamically stable. The stabilized K-doped phase $Ba_{1-x}K_xAsO_3$ ($x = 0.7$-$0.9$) is further predicted to exhibit phonon-mediated superconductivity. The corresponding phonon dispersion, the $\alpha^2F(\omega)$ and the $\lambda(\omega)$ are shown in Fig. S8. As shown in Fig S8, both acoustic branches and low-lying optical modes make non-negligible contributions to the EPC in $Ba_{1-x}K_xAsO_3$ ($x = 0.7$-$0.9$). Therefore, the REPME-based EPC renormalization scheme was not applied to these compounds, and the superconducting properties discussed here were based on the DFPT-LDA results. The calculated values of $\lambda$, $\omega_{log}$ and $T_c$ for these considered compounds are listed in Table S3. At $x = 0.7$, the $\lambda$ is 0.73, and the corresponding $T_c$ is about 12 K.

## V. CONCLUSION

In summary, we report two previously unexplored perovskites compound, $SrAsO_3$ and $BaAsO_3$, and systematically investigate their lattice instabilities, K-doping dynamical stabilization, and superconducting properties using first-principles calculations. The parent phases exhibit pronounced imaginary phonon modes, indicating strong dynamical instabilities. K-doping effectively suppresses these unstable phonon modes. With increasing K doping level, the imaginary phonon frequencies are progressively weakened, and no imaginary modes remain at sufficiently high doping levels. For $SrAsO_3$ and $BaAsO_3$, DFPT-LDA calculations indicate that dynamically stable cubic phases are obtained when the K doping level exceeds approximately 0.5 and 0.7, respectively.

We further analyze the electronic structures and superconducting properties of the K-doped phases. The stabilized cubic phases are metallic and with predicted conventional phonon mediated superconductivity. Within the DFPT-LDA framework, $T_c$ reaches approximately 36.1 K ($\lambda \sim$ 1.47) for $Sr_{1-x}K_xAsO_3$ at $x = 0.5$, and approximately 12 K ($\lambda \sim 0.73$) for $Ba_{1-x}K_xAsO_3$ at $x = 0.7$. For $Sr_{1-x}K_xAsO_3$, we further examined the influence of nonlocal exchange interactions by combining DFPT-LDA calculations with HSE06-based REPME corrections. The inclusion of nonlocal exchange effects substantially enhances the EPC strength across all investigated doping levels. For $Sr_{0.4}K_{0.6}AsO_3$, the $\lambda$ increases from 0.67 to 1.41, leading to a predicted $T_c$ of approximately 44.3 K.

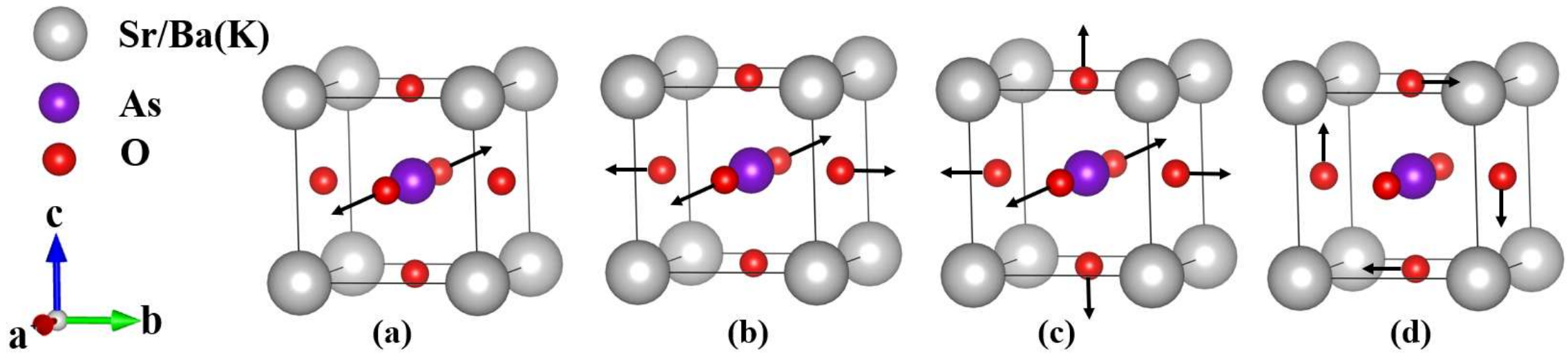

FIG. 1. Crystal structure of $A_{1-x}K_xAsO_3$(A = Ba, Sr) in the cubic perovskite phase. The arrows show (a) the oxygen-oscillating mode at the $X$ point (b) the oxygen-stretching mode at the $M$ point and (c) the oxygen-breathing mode at the $R$ point (d) the oxygen-rotational mode at the $M$ or $R$ point, respectively.

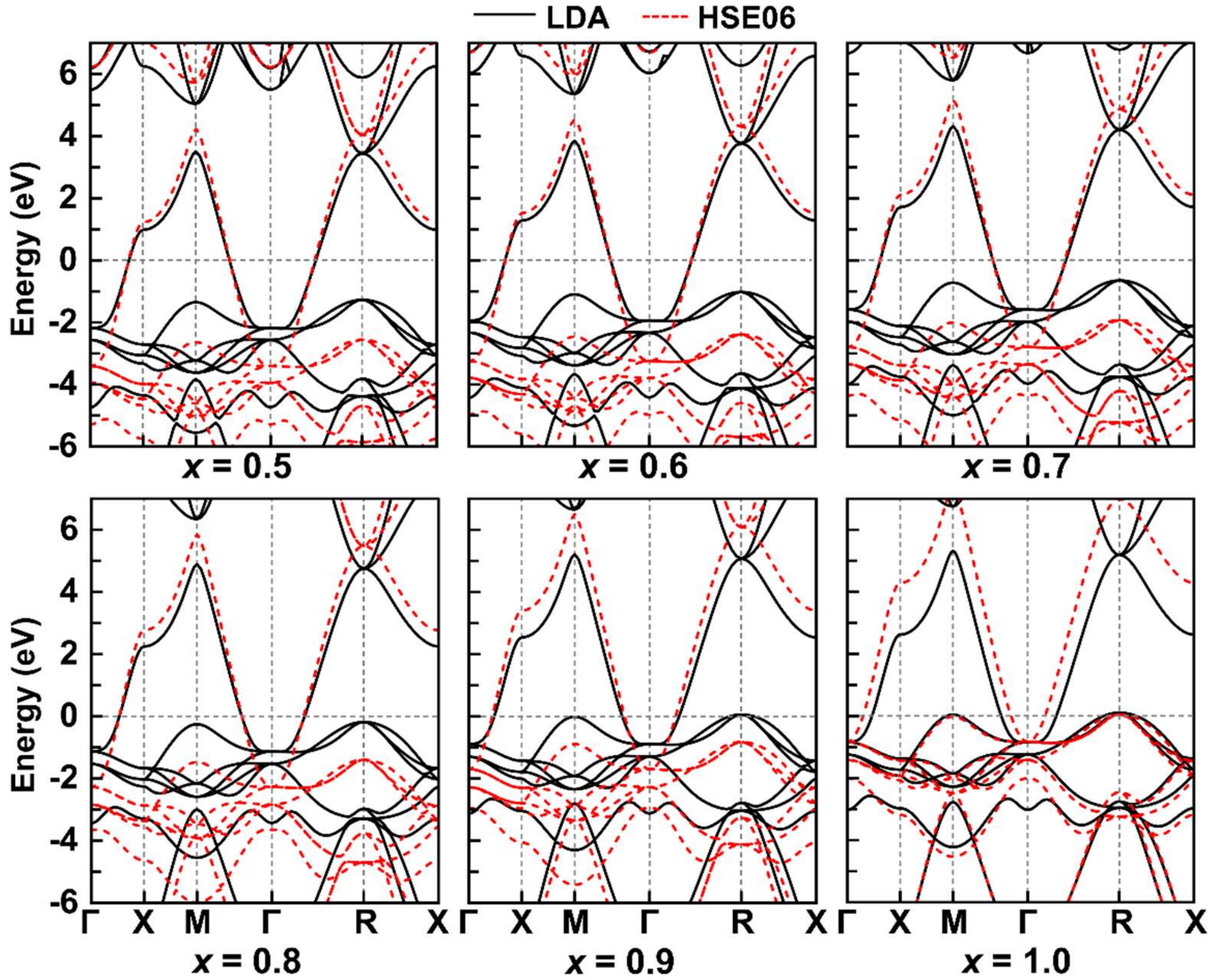

FIG. 2. Electronic band structures of $Sr_{1-x}K_xAsO_3$ ($x$ = 0.5, 0.6, 0.7, 0.8, 0.9, 1.0) by using both the LDA and HSE06 hybrid functionals.

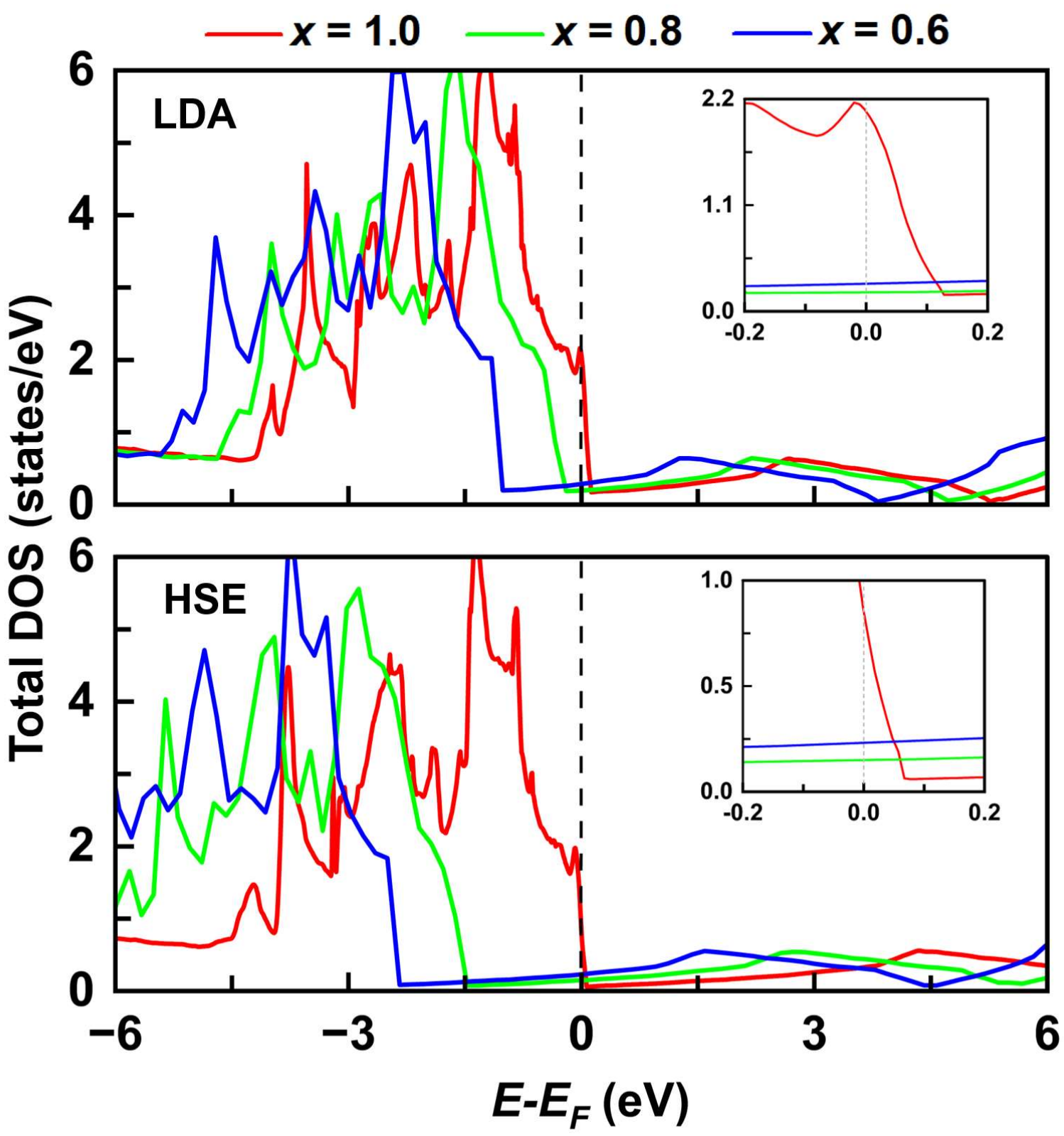

FIG. 3. Total electronic density of states of $Sr_{1-x}K_xAsO_3$ ($x$ = 0.6, 0.8, 1.0) by using both (a) LDA and (b) HSE06 hybrid functionals. The insets show the electronic DOS near the Fermi energy.

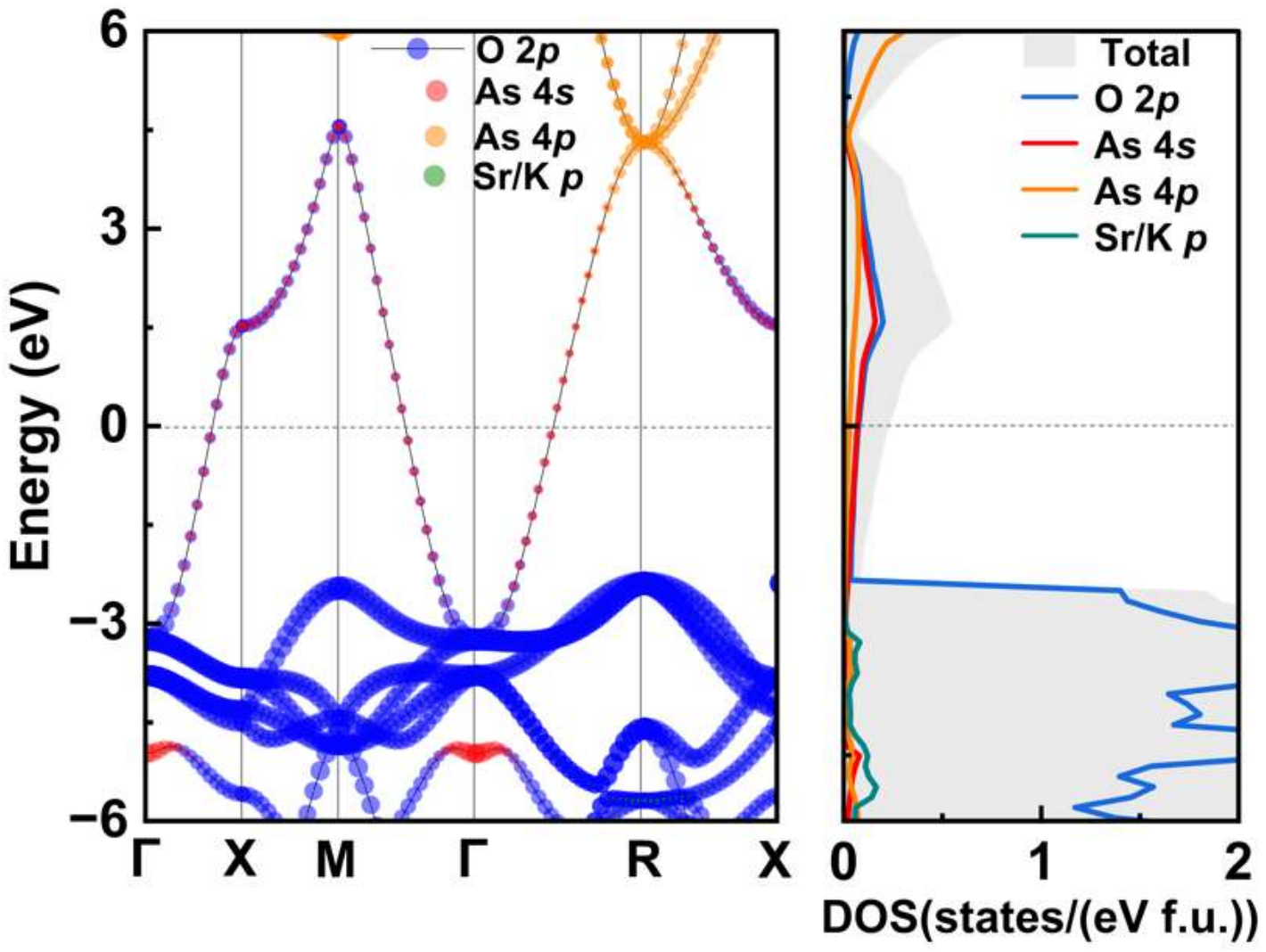

FIG. 4. Electronic band structure (left) and density of states (right) of $Sr_{0.4}K_{0.6}AsO_3$ using the HSE06 hybrid functionals

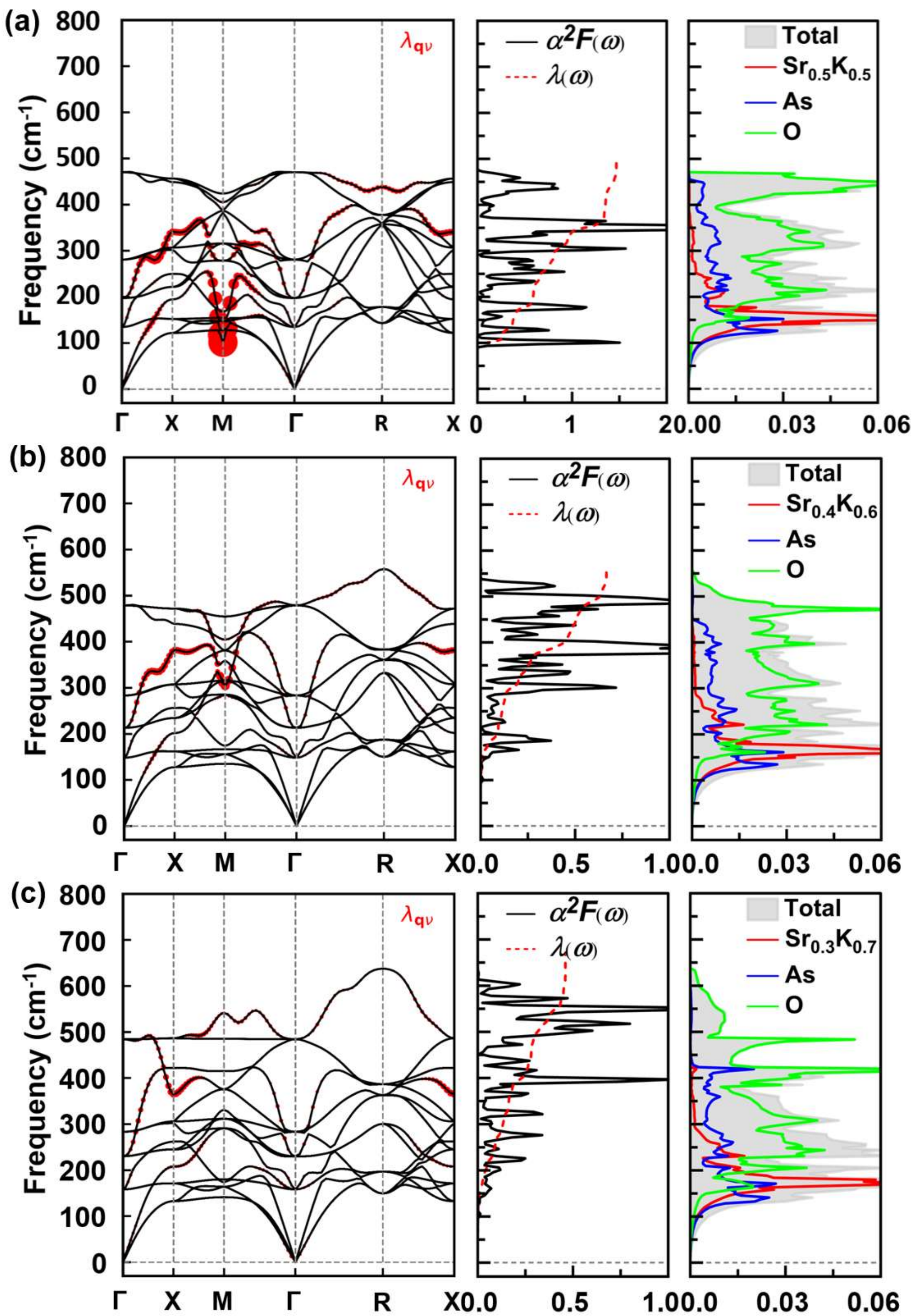

(a)
(b)
(c)
Frequency (cm-1)
800
700
600
500
400
300
200
100
0
Γ X M Γ R X
λqν
α2F(ω)
λ(ω)
Total
Sr0.5K0.5
Sr0.4K0.6
Sr0.3K0.7
As
O
0 1 2
0.0 0.5 1.0
0.00 0.03 0.06

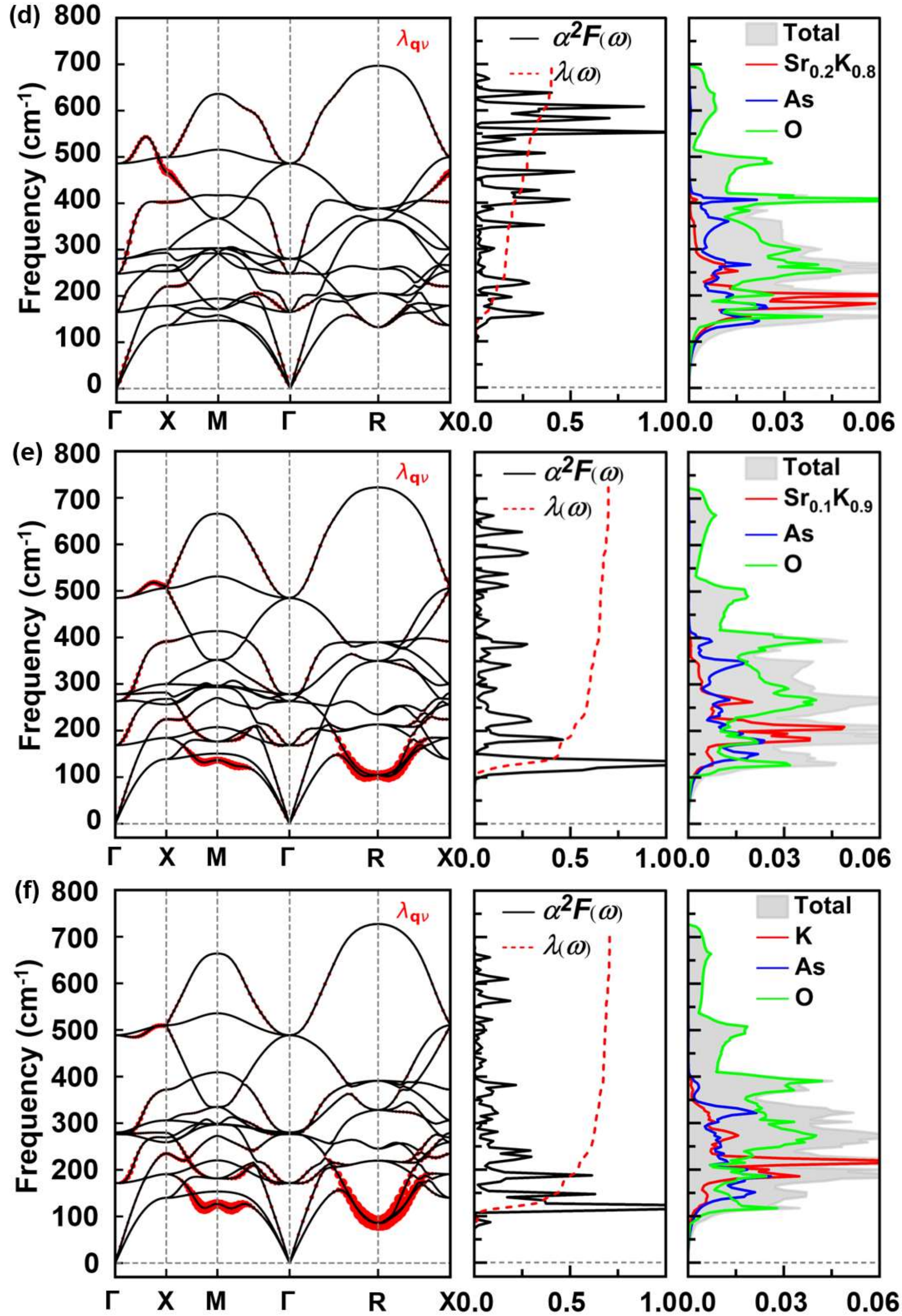

FIG. 5. The DFPT-LDA calculated phonon spectra (left panel, the radius of the red dots is proportional to $\lambda_{q\nu}$), Eliashberg function $\alpha^2F(\omega)$ (middle panel), and phonon density of states (right panel) of $Sr_{1-x}K_xAsO_3$. (a) $x = 0.5$, (b) $x = 0.6$, (c) $x = 0.7$, (d) $x = 0.8$, (e) $x = 0.9$, (f) $x = 1.0$

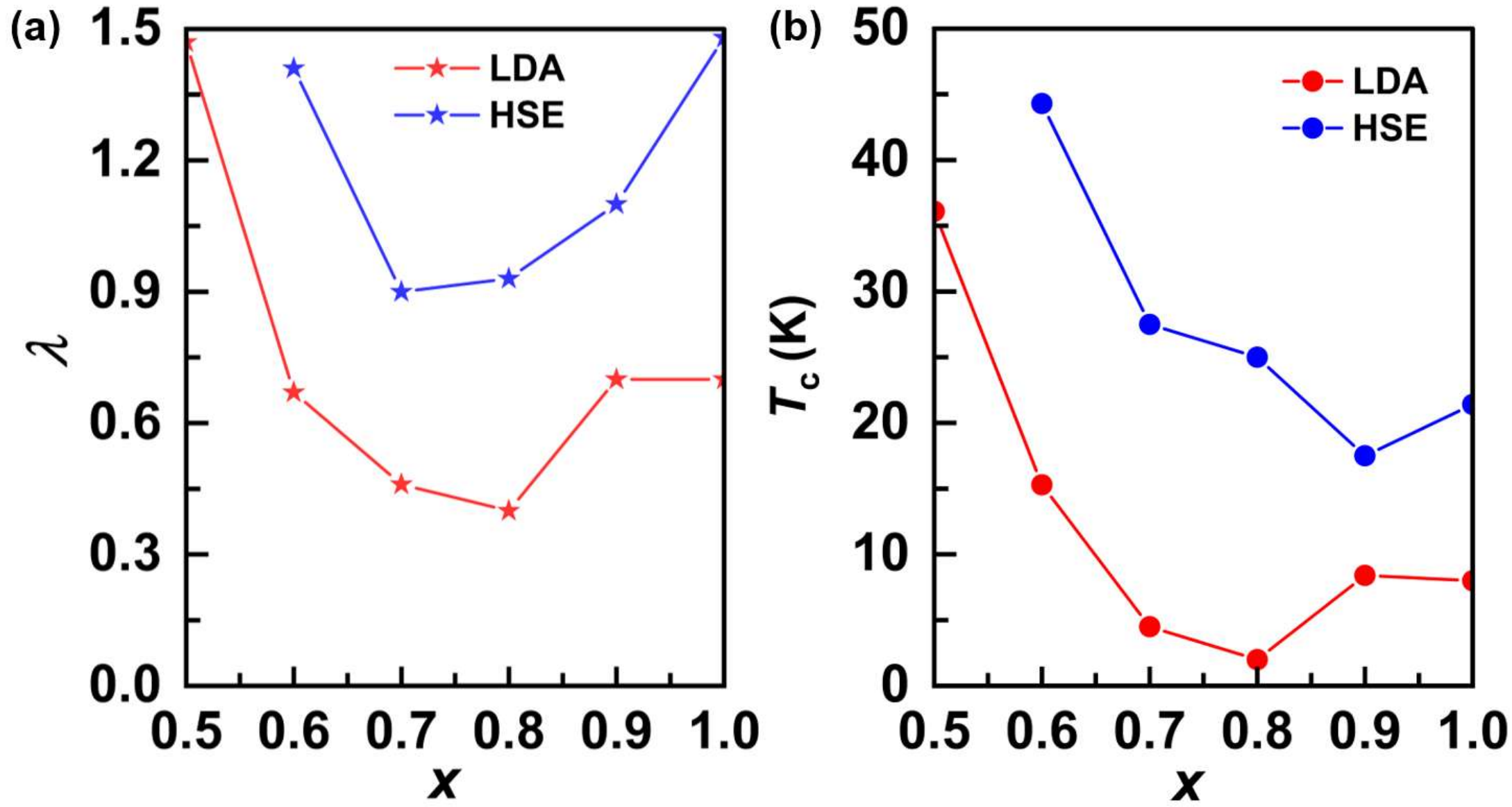

FIG. 6. The (a) total EPC $\lambda$, and (b) $T_c$ (K) calculated by the LDA and the HSE06 hybrid functionals for $Sr_{1-x}K_xAsO_3$($x$ = 0.5, 0.6, 0.7, 0.8, 0.9, 1.0).

TABLE 1. The total EPC $\lambda$, the average phonon frequency $\omega_{\log}$ (K), and the calculated $T_c$ (K) in the DFPT-LDA of $Sr_{1-x}K_xAsO_3$($x$ = 0.5, 0.6, 0.7, 0.8, 0.9, 1.0) . (For $x$ = 0.9, only 59% of the $\lambda$ was renormalized within the HSE06-based correction scheme, the rationale for this treatment is provided in the Supplementary Material)

| | LDA | | | $\left\langle \frac{\lvert D_H^\nu \rvert^2}{\lvert D_L^\nu \rvert^2} \right\rangle$ | HSE | | |
|---|---|---|---|---|---|---|---|
| $x$ | $\lambda$ | $\omega_{\log}$ | $T_c$ | | $\lambda$ | $\omega_{\log}$ | $T_c$ |
| 1.0 | 0.70 | 230.2 | 8.0 | 2.11 | 1.48 | 190.6 | 21.4 |
| 0.9 | 0.70 | 243.5 | 8.4 | 1.97 | 1.10 | 219.1 | 17.5 |
| 0.8 | 0.40 | 472.4 | 2.0 | 2.33 | 0.93 | 402.0 | 25.0 |
| 0.7 | 0.46 | 531.0 | 4.5 | 1.97 | 0.90 | 464.9 | 27.5 |
| 0.6 | 0.67 | 497.7 | 15.3 | 2.11 | 1.41 | 413.9 | 44.3 |
| 0.5 | 1.47 | 324.1 | 36.1 | | | | |

## ACKNOWLEDGMENTS

This work was supported by the National Natural Science Foundation of China (Grant No. 12574155, 12074041 and 12404160), the Fundamental Research Funds for the Central Universities (Grant No. 2243300003), and Fundamental Research Program of Shanxi Province (202203021222228). The calculations were carried out with high performance computing cluster of Beijing Normal University in Zhuhai.

.

## DATA AVAILABILITY

The data that support the findings of this article are not publicly available. The data are available from the authors upon reasonable request.

*Requests for materials should be addressed to Z.P.Y. at yinzhiping@bnu.edu.cn

**Supplemental Materials for:**

**Prediction of $BaBiO_3$-like superconducting perovskites in K-doped $SrAsO_3$**

For $x = 0.9$ and $x = 1.0$, the DFPT-LDA calculations show that the oxygen-rotational modes at the $R$ and $M$ points provide a major contribution to the EPC (Fig 5(e) and 5(f)).

For $x = 0.9$, the electronic structures obtained from the two functionals exhibit an important difference. As discussed in Sec. 3.2, the DFT-LDA band structure (Fig. 2) shows that a valence band crosses the Fermi level, whereas in the DFT-HSE06 calculations the corresponding band lies below the Fermi level. To assess whether the EPC contributions associated with the rotational modes should be retained in the HSE06-based renormalization, frozen phonon calculations were performed within the DFT-LDA framework. As shown in Figs. S5(c) and S5(d), the largest phonon induced band splittings occur on the valence band crossing the Fermi level. This indicates that the strong EPC associated with the $R$ and $M$ point rotational modes is closely related to the electronic states of this band. In the HSE06 calculations, however, this valence band is shifted below the Fermi level. Therefore, the EPC contributions associated with the low frequency oxygen-rotational branches connected to the $M$- and $R$-point rotational modes were excluded from the HSE06-based renormalization. For $x = 0.9$, according to the DFPT-LDA results, these modes account for approximately 41% of the total EPC strength. Consequently, only the remaining 59% of the EPC strength was renormalized using the REPME based correction scheme.

For $x = 1.0$, both DFT-LDA and DFT-HSE06 calculations show that the relevant valence band crosses the Fermi level, in contrast to the $x = 0.9$ case. Frozen phonon calculations further confirm that the phonon-induced band splittings associated with the *R* and *M* point oxygen-rotational modes occur predominantly on the valence band crossing the Fermi level. Therefore, the oxygen-rotational modes at the *R* and *M* points were also included in the REPME analysis, together with the oxygen oscillating mode at the *X* point. The corresponding frozen phonon band structures are shown in Figs. S6(b) and S6(c), while the extracted band splittings and REPMEs are summarized in Table S3.

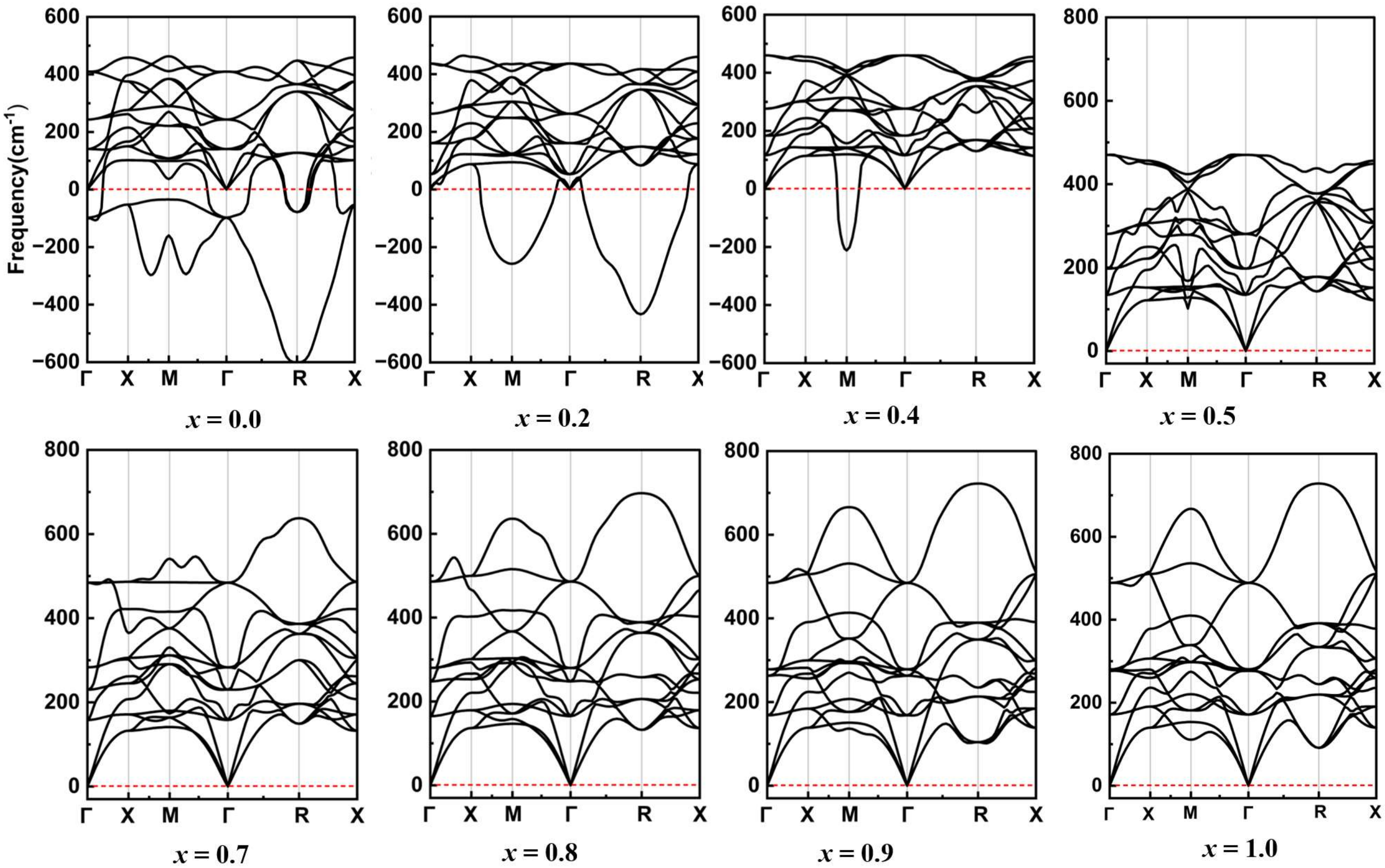

FIG. S1. The DFPT-LDA calculated phonon spectra of $Sr_{1-x}K_xAsO_3$ ($x$ = 0-1.0)

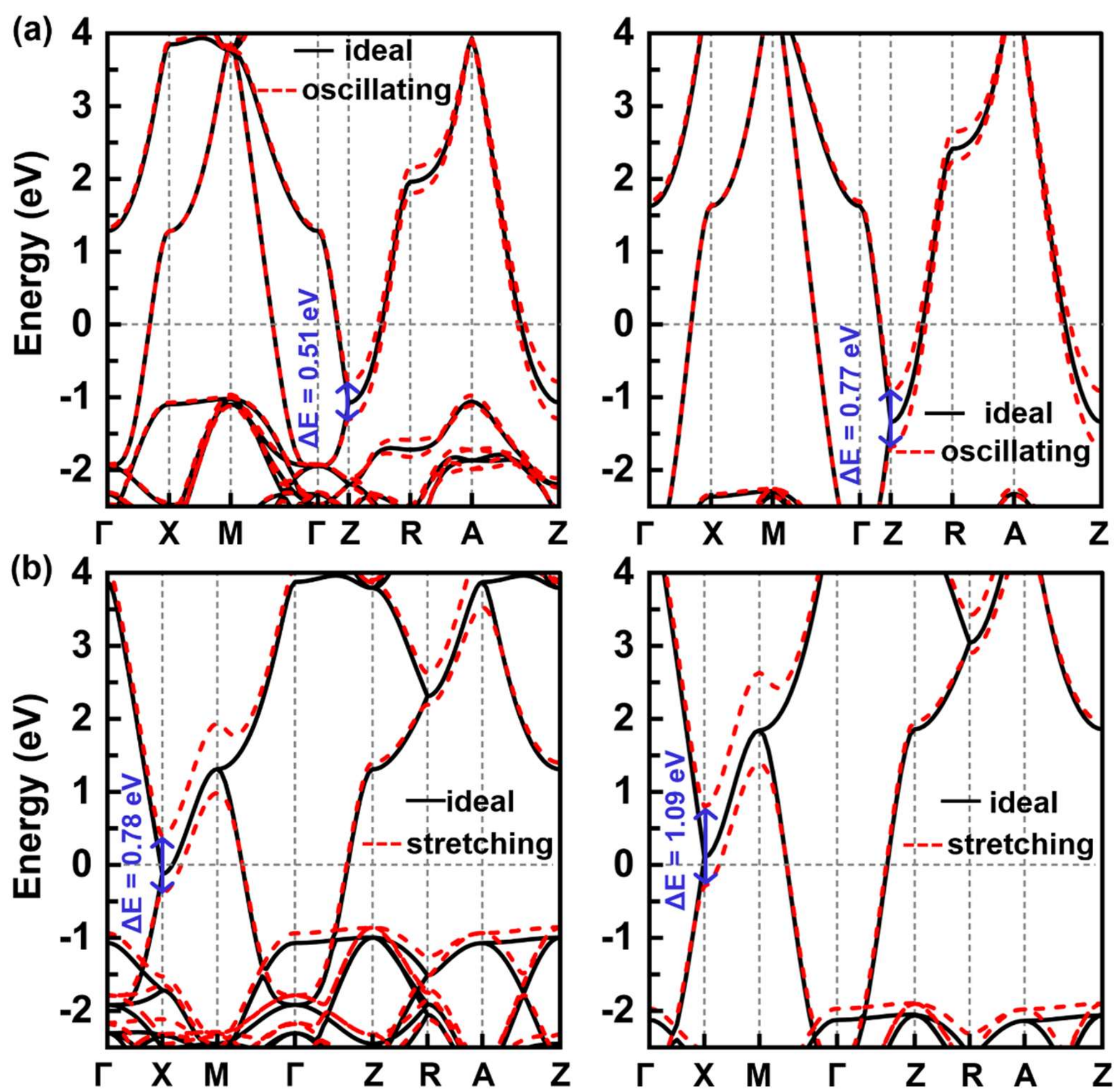

FIG. S2. Illustration of the REPMEs of (a) oxygen-oscillating mode at the *X* point, (b) oxygen-stretching mode at the *M* point in $Sr_{0.4}K_{0.6}AsO_3$. The band structures with and without the oxygen displacement are calculated by DFT using both LDA (left panel) and HSE06 (right panel) hybrid functionals.

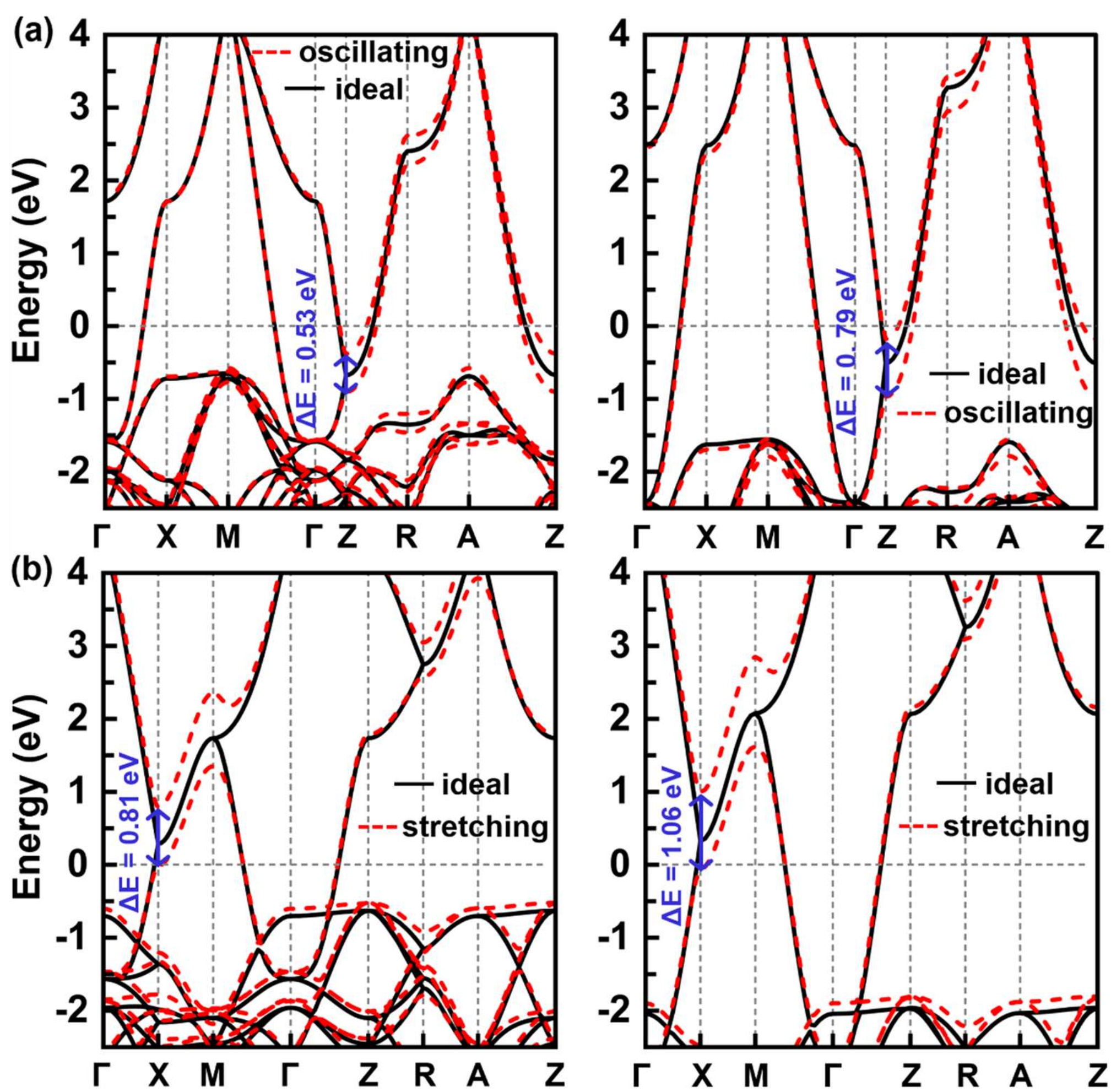

FIG. S3. Illustration of the REPMEs of (a) oxygen-oscillating mode at the $X$ point, (b) oxygen-stretching mode at the $M$ point in $Sr_{0.3}K_{0.7}AsO_3$. The band structures with and without the oxygen displacement are calculated by DFT using both LDA (left panel) and HSE06 (right panel) hybrid functionals.

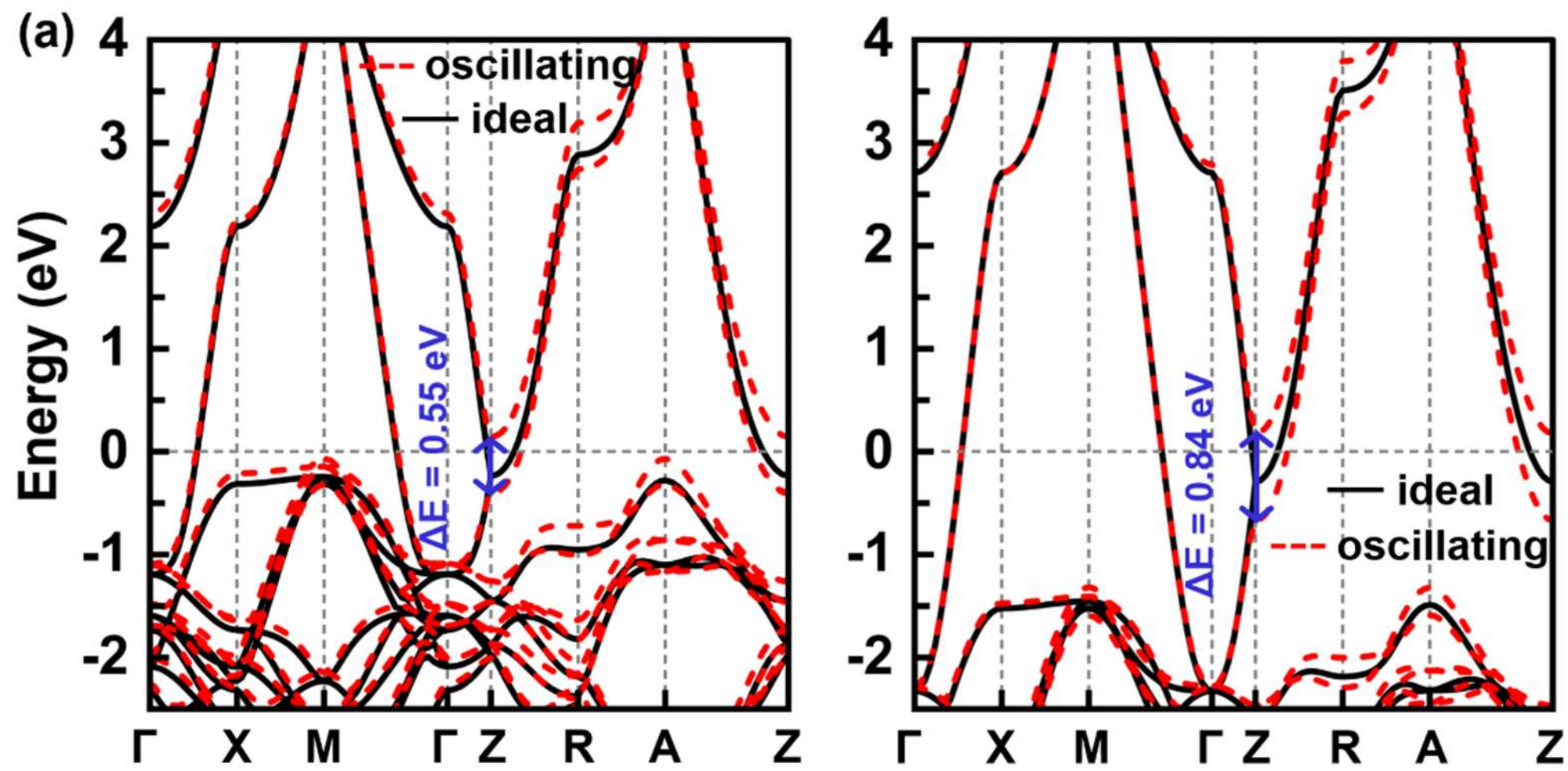

FIG. S4. Illustration of the REPMEs of (a) oxygen-oscillating mode at the *X* point in $Sr_{0.2}K_{0.8}AsO_3$. The band structures with and without the oxygen displacement are calculated by DFT using both LDA (left panel) and HSE06 (right panel) hybrid functionals.

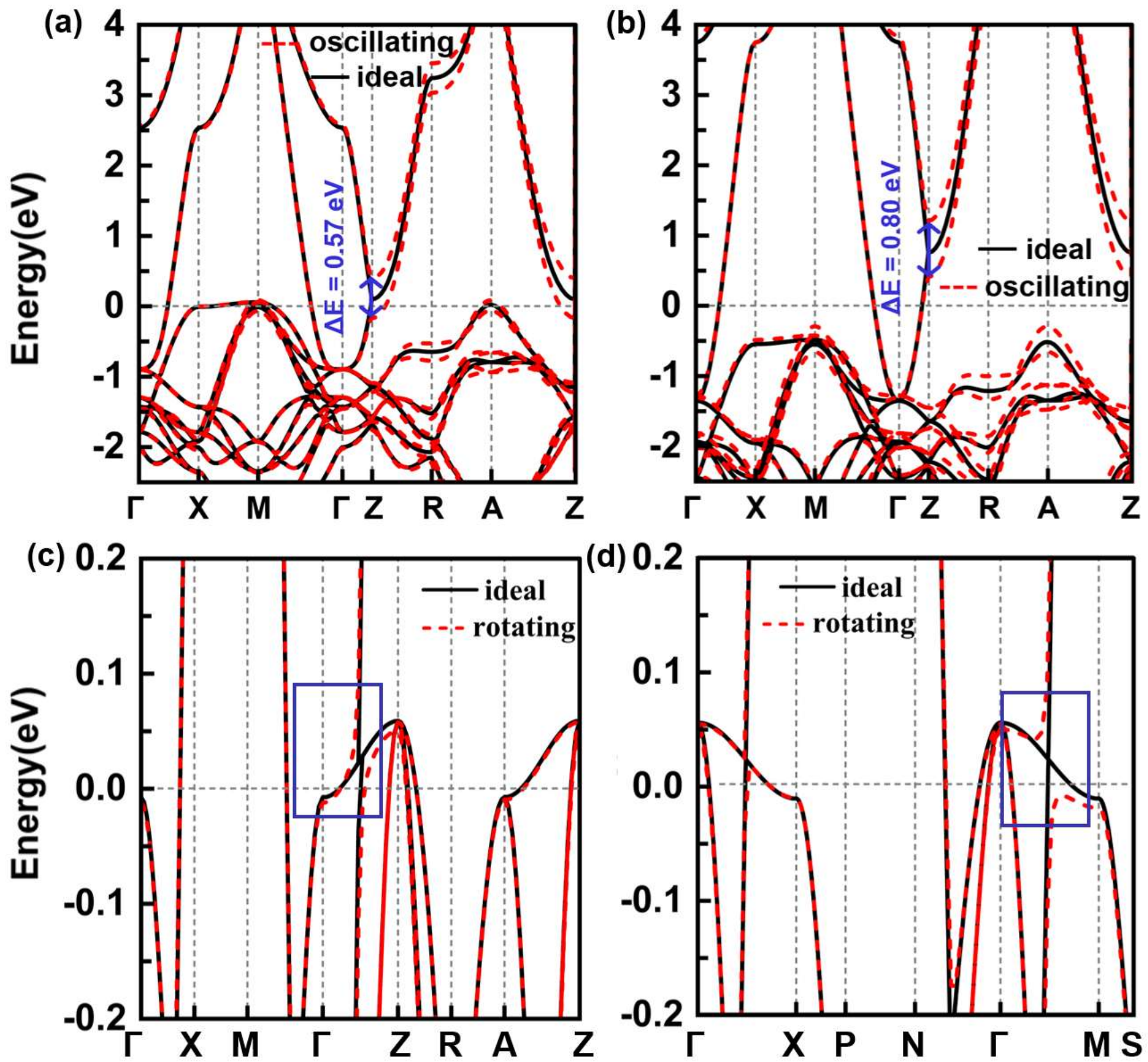

FIG. S5. Illustration of the REPMEs of (a) oxygen-oscillating mode at the *X* point (The band structures with and without the oxygen displacement are calculated by DFT-LDA), b) oxygen-oscillating mode at the *X* point (The band structures with and without the oxygen displacement are calculated by DFT-HSE06), (c) Oxygen-rotating mode at the *M* point (the band structures with and without the oxygen displacement are calculated using DFT-LDA), (d) Oxygen-rotating mode at the *R* point (the band structures with and without the oxygen displacement are calculated using DFT-LDA) in $Sr_{0.1}K_{0.9}AsO_3$.

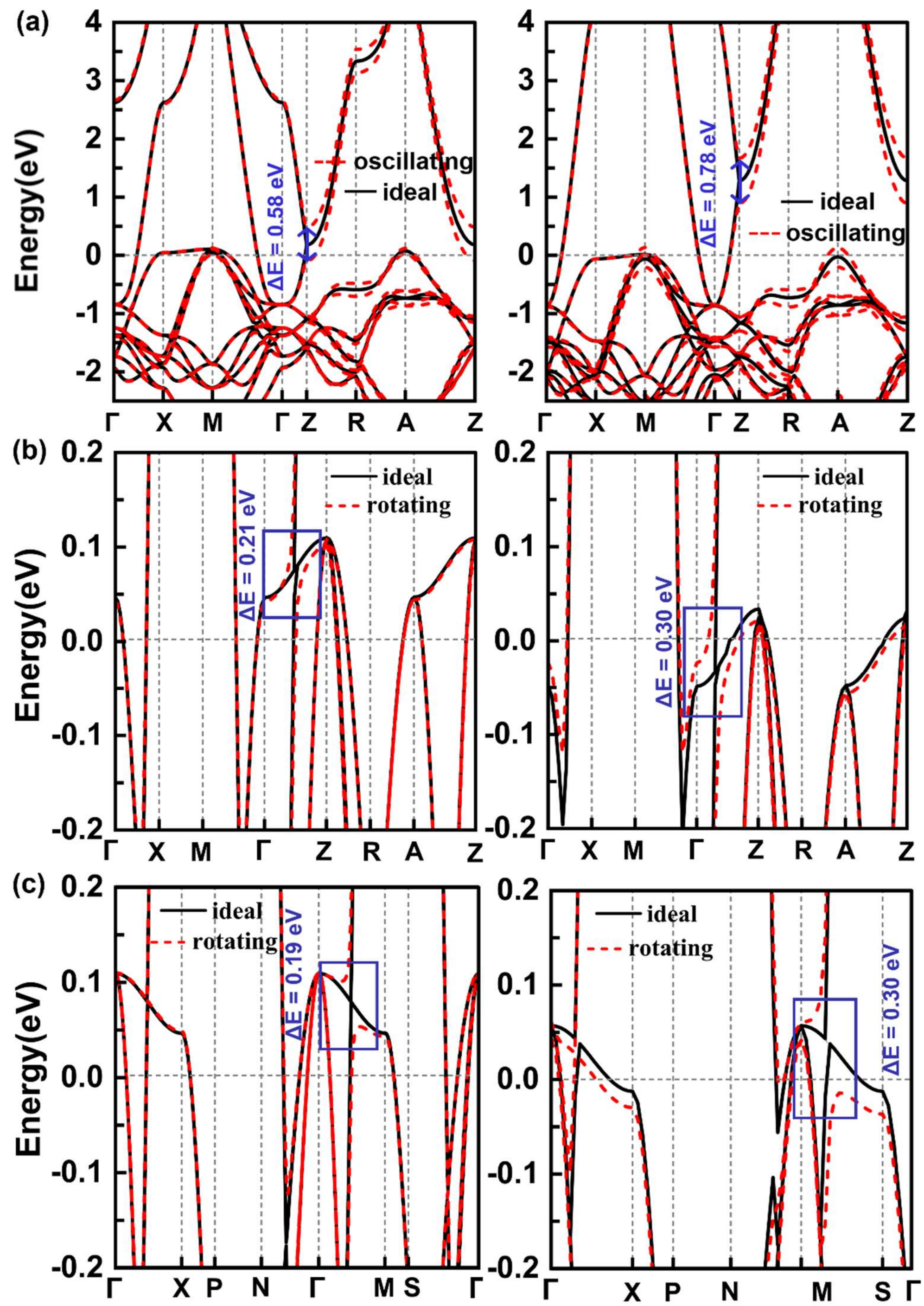

FIG. S6. Illustration of the REPMEs of (a) oxygen-oscillating mode at the *X* point, (b) oxygen-rotating mode at the *M* point, (c) oxygen-rotating mode at the *R* point in $KAsO_3$. The band structures with and without the oxygen displacement are calculated by DFT using both LDA (left panel) and HSE06 (right panel) hybrid functionals.

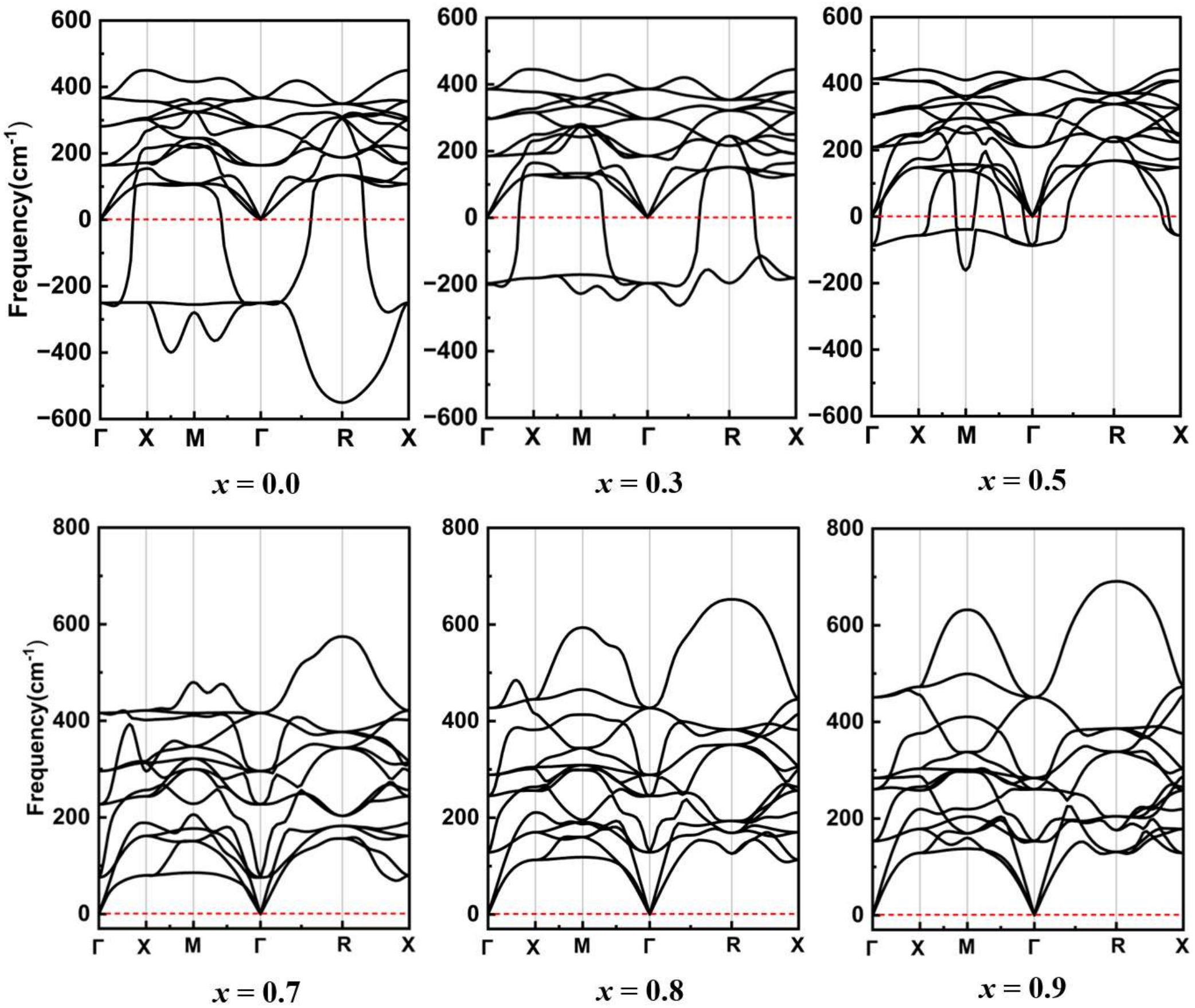

FIG. S7. The DFPT-LDA calculated phonon spectra of $Ba_{1-x}K_xAsO_3$ ($x$ = 0-0.9)

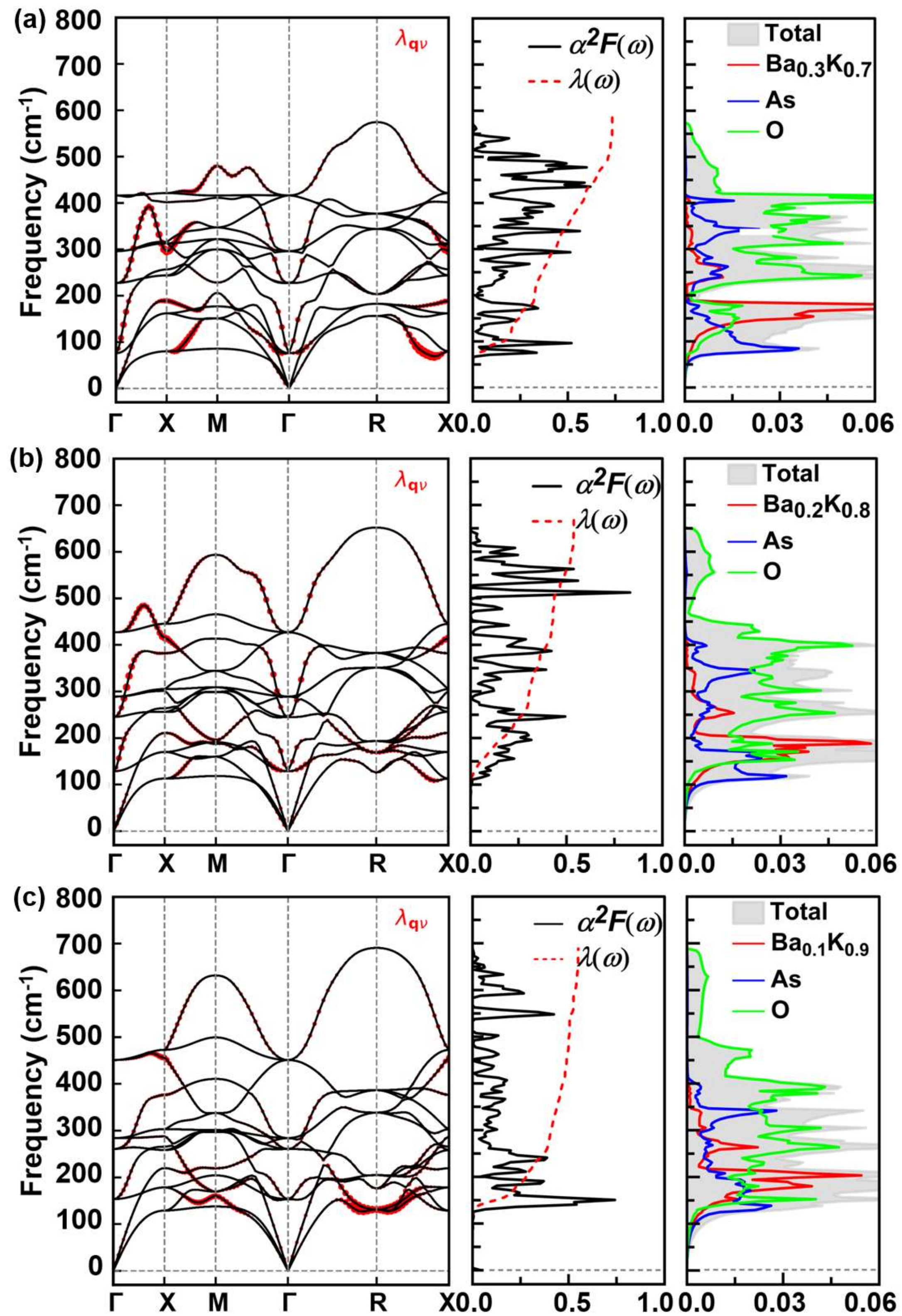

FIG. S8. The DFPT-LDA calculated phonon spectra (left panel, the radius of the red dots is proportional to $\lambda_{qv}$), Eliashberg function $\alpha^2F(\omega)$ (middle panel), and phonon density of states (right panel) of $Ba_{1-x}K_xAsO_3$. (a) $x$ = 0.7, (b) $x$ = 0.8, (c) $x$ = 0.9

TABLE S1. The bandwidths (eV) along Γ-X direction and Γ-M direction in the LDA and the HSE06 hybrid functionals for $Sr_{1-x}K_xAsO_3$ ($x$ = 0.5, 0.6, 0.7, 0.8, 0.9, 1.0).

| $x$ | Γ-X | | | Γ-M | | |
|---|---|---|---|---|---|---|
| | LDA | HSE06 | Broadening | LDA | HSE06 | Broadening |
| 0.5 | 3.16 | 4.64 | 47% | 5.69 | 7.63 | 34% |
| 0.6 | 3.23 | 4.77 | 48% | 5.79 | 7.79 | 35% |
| 0.7 | 3.30 | 4.90 | 49% | 5.90 | 7.96 | 35% |
| 0.8 | 3.38 | 5.04 | 49% | 6.02 | 8.12 | 35% |
| 0.9 | 3.43 | 5.10 | 49% | 6.10 | 8.20 | 34% |
| 1.0 | 3.47 | 5.12 | 48% | 6.15 | 8.22 | 34% |

TABLE S2. The band splittings (eV), and REPMEs $D$ (eV/Å) for the most important vibration modes in the LDA and the HSE06 hybrid functionals for $Sr_{1-x}K_xAsO_3$ ($x$ =0.6, 0.7, 0.8, 0.9,1.0).

| | | Band splitting | | $D_L$ | $D_H$ | $\left\langle \frac{\lvert D_H^\nu \rvert^2}{\lvert D_L^\nu \rvert^2} \right\rangle$ |
|---|---|---|---|---|---|---|
| | | LDA | HSE | LDA | HSE | |
| O-stretching at M | 0.6 | 0.78 | 1.09 | 10.32 | 14.42 | 1.95 |
| | 0.7 | 0.81 | 1.06 | 10.77 | 14.10 | 1.71 |
| O-oscillating at X | 0.6 | 0.51 | 0.77 | 6.75 | 10.19 | 2.28 |
| | 0.7 | 0.53 | 0.79 | 7.05 | 10.51 | 2.22 |
| | 0.8 | 0.55 | 0.84 | 7.35 | 11.22 | 2.33 |
| | 0.9 | 0.57 | 0.80 | 7.64 | 10.73 | 1.97 |
| | 1.0 | 0.58 | 0.78 | 7.79 | 10.48 | 1.81 |
| O-rotational at M | 1.0 | 0.21 | 0.30 | 2.82 | 4.03 | 2.04 |
| O-rotational at R | 1.0 | 0.19 | 0.30 | 2.55 | 4.03 | 2.49 |

TABLE S3. The total EPC $\lambda$, the average phonon frequency $\omega_{\log}$ (K), and the calculated $T_c$ (K) in the DFPT-LDA of $Ba_{1-x}K_xAsO_3$($x$ = 0.7, 0.8, 0.9)

| $x$ | $\lambda$ | $\omega_{\log}$ | $T_c$ |
|---|---|---|---|
| 0.9 | 0.56 | 314.7 | 5.9 |
| 0.8 | 0.54 | 394.0 | 6.3 |
| 0.7 | 0.73 | 308.5 | 12.0 |